\documentclass[12pt]{iopart}
\usepackage{amsmath}
\usepackage{iopams}  
\usepackage{graphicx}
\usepackage{dcolumn}
\usepackage{bm}
\usepackage{float}
\usepackage{hyperref}
\usepackage{xcolor}

\def\endproof{\vrule height6pt width6pt depth0pt}

\newcommand{\mean}[1]{\left<#1\right>}

\newtheorem{result}{Result}

\newtheorem{conjecture}{Conjecture}

\newtheorem{lemma}{Lemma}

\begin{document}

\title[Trade-off relations between Bell nonlocality and local Kochen-Specker contextuality]{Trade-off relations between Bell nonlocality and local Kochen-Specker contextuality in generalized Bell scenarios}

\author{Lucas E. A. Porto $^1$ $^2$, Gabriel Ruffolo $^1$ $^3$, Rafael Rabelo $^1$, Marcelo {Terra Cunha} $^4$, Pawe\l{} Kurzy\ifmmode \acute{n}\else \'{n}\fi{}ski $^3$}

\address{$^1$ Instituto de F\'{i}sica Gleb Wataghin, Universidade Estadual de Campinas (Unicamp),
Rua S\'{e}rgio Buarque de Holanda 777, Campinas, S\~{a}o Paulo 13083-859, Brazil}
\address{$^2$ Sorbonne Université, CNRS, LIP6, Paris F-75005, France}
\address{$^3$ Institute of Spintronics and Quantum Information, Faculty of Physics, Adam Mickiewicz University, 61-614 Pozna\'n, Poland}
\address{$^4$ Instituto de Matem\'{a}tica, Estat\'{i}stica e Computa\c{c}\~{a}o Cient\'{i}fica, Universidade Estadual de Campinas (Unicamp), Rua S\'{e}rgio Buarque de Holanda 651, Campinas, S\~{a}o Paulo 13083-859, Brazil}

\ead{lporto@ifi.unicamp.br}

\begin{indented}
\item[]\today
\end{indented}

\begin{abstract}
The relations between Bell nonlocality and Kochen-Specker contextuality have been subject of research from many different perspectives in the last decades. 
Recently, some interesting results on these relations have been explored in the so-called generalized Bell scenarios, that is, scenarios where Bell spatial separation (or agency independence) coexist with (at least one of the) parties' ability to perform compatible measurements at each round of the experiment. 
When this party has an $n$-cycle compatiblity setup, it was first claimed that Bell nonlocality could not be concomitantly observed with contextuality at this party's local experiment. However, by a more natural reading of the definition of locality, it turns out that both Bell nonlocality and local contextuality can, in fact, be jointly present. 
In spite of it, in this work we prove that in the simplest of those scenarios there cannot be arbitrary amounts of both of these two resources together. 
That is, in these cases we show that the violation of any Bell inequality limits the possible violations of any local noncontextuality inequality. We also explore this trade-off relation using quantifiers of nonlocality and contextuality, discussing how such a relation can be understood in terms of a `global' notion of contextuality, and we study possible extensions of this result to other scenarios.
\end{abstract}

%
%
%
\maketitle
%
%

\section{Introduction}
Bell nonlocality \cite{Brunner2014, Bell1964} and Kochen-Specker contextuality \cite{Kochen1967, Budroni2021} are two of the most remarkable features of quantum theory. 
They reveal a fundamental contrast between its predictions and those of classical theories. 
As such, they constitute some of the cornerstones for how our current understanding of quantum theory is conceived.
They are also essential resources for quantum advantages in informational tasks \cite{ekert91, howard2014, raussendorf2013}. 

Nonlocality and contextuality are in fact closely related concepts, since they both are associated to stronger-than-classical correlations between the outcomes of compatible measurements. 
Bell nonlocality can actually be seen as a particular case of contextuality, which takes place in scenarios where there exist multiple observers far apart (or informationally isolated) from each other and the compatibility relations are solely inherited from such spatial separation.

The similarity between these two concepts motivates investigations from many different perspectives to better understand their connections \cite{scala2024, soeda2013, hu2018, saha2017}. 
For example, since the first decades after they were discovered, there is an ongoing search for how to convert proofs of contextuality into proofs of nonlocality \cite{stairs1983, heywood1983, cabello2021, Cabello2020}. 
In another direction, the Cabello-Severini-Winter graph-theoretic approach \cite{CSW2010, CSW2014} can be adapted to demand compatibilities coming from separated parties \cite{Rabelo14}, allowing one to compare the sets of quantum correlations coming from Bell or Bell-like scenarios with the larger quantum sets allowed in contextuality scenarios \cite{Vandre22, porto23}.

In yet another way, the relations between nonlocality and contextuality can be explored in the framework of the so-called \textit{generalized Bell scenarios} \cite{Mazzari23, Kurzynski2014}. 
These are Bell scenarios where (at least) one of the parties is allowed to perform compatible measurements on their local system at each round of the experiment. 
This party is then able to make a local contextuality test at the same time as the nonlocality experiment is performed amongst all the observers \cite{xue2023}. 
These are precisely the scenarios studied in this work.

In recent years, some rather surprising results on the simplest generalized Bell scenarios have been explored in literature. 
For example, in the bipartite scenario where Alice has two incompatible measurements and Bob has a $5$-cycle setup \cite{Araujo2013}, local contextuality on Bob's experiment has been observed concomitantly with Bell nonlocal correlations between him and Alice \cite{xue2023}, a phenomenon that was previously thought to be impossible \cite{Kurzynski2014, jia2016}. The key fact behind this joint observation is a suitable modification in the definition of Bell nonlocality, specifically tailored for the generalized scenarios \cite{Temistocles2019, Mazzari23}. This new definition is in fact more powerful than the usual one, and leads to new relevant classes of Bell-like inequalities. One of these new inequalities, in particular, is then used to jointly witness nonlocality and contextuality in the above mentioned scenario \cite{xue2023}.

In this work, we go deeper in this investigation and analyse the question of whether it is possible to jointly observe \textit{arbitrary} amounts of nonlocality and contextuality in generalized Bell scenarios. In the scenarios where Alice has two incompatible measurements and Bob has a 3, 4 or 5-cycle, we show that this is not the case, by proving that the sum of the violations of any given Bell inequality and any given noncontextuality inequality has a nontrivial bound, thus characterizing a trade-off relation between these two features. Moreover, by rewriting this result in terms of quantifiers of nonlocality and contextuality, we show how it can be understood as coming from a global notion of contextuality for generalized Bell scenarios. Finally, we study the existence of this trade-off relation in other scenarios, providing evidences that it holds if Bob has an arbitrary $n$-cycle, but also showing one example of a scenario where it does not hold.

The manuscript is structured as follows. 
In Section \ref{sec:preliminaries}, we discuss some of the background required and introduce the concepts related to generalized Bell scenarios. 
The presentation of our results starts in Section \ref{sec:trade-off}, where we show a trade-off relation between Bell nonlocality and local contextuality holding in scenarios where Alice performs two binary measurements and Bob performs an $n$-cycle with small $n$, and discuss some consequences of such a relation. 
In section \ref{sec:trade-off-quantifiers}, we present a version of this trade-off based on quantifiers, which includes a more general concept of nonclassicality that can be defined in generalized Bell scenarios. 
In section \ref{sec:more_scenarios} we explore other scenarios where trade-off relations could appear and we give an example where they do not hold.  We conclude in Section \ref{sec:conclusion}, also discussing some of the further questions raised by this work.

\section{Preliminaries}\label{sec:preliminaries}
In this section, we present some of the definitions and previous results relevant to our discussion. In particular, we introduce the generalized Bell scenarios and discuss some of the distinct definitions of locality applicable to them \cite{Temistocles2019, Mazzari23}. Also, we present some of the recent results relating nonlocality and local contextuality on these scenarios.

\subsection{Basic notions of Kochen-Specker contextuality}\label{sec:contextuality_basics}

Let us recall some of the basic concepts related to Kochen-Specker contextuality \cite{Budroni2021}. A \textit{contextuality scenario} is specified by a set of measurements $\mathcal{M}$, their compatibility relations, and a set of possible outcomes for each of them. The compatibility relations are typically specified by a set $\mathcal{C}$, whose elements are subsets of $\mathcal{M}$ composed by compatible observables. The elements of $\mathcal{C}$, that is, the sets of measurements which can be jointly performed, are named \textit{contexts}.

To describe a contextuality experiment we need, for each context $C \in \mathcal{C}$, a probability distribution $p_C$ for its possible outcomes. These probabilities, considering all contexts and all possible outcomes, can be organized as the entries of a vector $\boldsymbol{p} \in \mathbb{R}^d$ in an appropriate dimension $d$. This vector is called \textit{behavior} of the experiment.

We require the acceptable behaviors in a contextuality scenario to satisfy some consistency conditions, known as \textit{no-disturbance} conditions, stating that marginals be well defined. More specifically, for every $C, C' \in \mathcal{C}$, these conditions can be written as
\begin{equation}
    p_{C\vert C \cap C'} = p_{C'\vert C \cap C'},
    \label{eq:no-disturbance}
\end{equation}
where $p_{C\vert C \cap C'}$ denotes the restriction of the distribution $p_C$ to the measurements in the intersection $C\cap C'$.

A no-disturbance behavior is said to be \textit{noncontextual} if there exists a variable $\lambda$, described by a probability distribution $q(\lambda)$, such that for every $C \in \mathcal{C}$ we can write
\begin{equation}\label{eq:noncontextuality}
    p_C(\boldsymbol{s}) = \sum_\lambda q(\lambda) \prod_{M \in C}p_{M}^\lambda (s_M),
\end{equation}
where $\boldsymbol{s}$ denotes the outcome of the context $C$, and $s_M$ denotes the corresponding outcome of the measurement $M \in C$. Moreover, $p_{M}^\lambda$ is a probability distribution on the outcomes of the measurement $M$ which may depend on $\lambda$.  

Fine's theorem states that the distributions $p^\lambda_M$ might be taken to be deterministic without loss of generality \cite{Fine82}. Thus, the noncontextual behaviors are convex combinations of deterministic behaviors, that is, behaviors whose entries are either $0$ or $1$, respecting the no-disturbance conditions. So, in a noncontextual behavior it is possible to think of measurements as if they were simply revealing a property of the system which was already well defined prior to the measurements. In this case, the probabilities only arise as a consequence of a lack of knowledge of the system's preparation, in accordance with what we would expect within a classical theory.  

Geometrically, both the set of nondisturbing behaviors and the set of noncontextual behaviors in $\mathbb{R}^d$ are \textit{polytopes}. In a typical contextuality scenario, we easily know how to specify the \emph{noncontextual polytope} by its vertices, which are the deterministic behaviors. However, for many reasons it is desirable to characterize it in terms of inequalities, often called \textit{noncontextuality inequalities}. On the other hand, we easily know how to specify the \emph{nondisturbing polytope}, which contains the noncontextual polytope, via the inequality description using no-disturbance conditions and non-negativity and normalization of probabilities. Sometimes, however, it is useful to characterize it via the vertex description. Changing between these two descriptions, for both the noncontextual and the nondisturbing polytopes, is typically a hard task, and has only been done analytically in a few cases, such as the $n$-cycle scenarios \cite{Araujo2013}. 

\subsection{Usual and generalized Bell scenarios}\label{sec:bell_scenarios} 
Usually, to define a Bell scenario we specify the number of measurements each of the parties can perform on her respective subsystem, as well as the number of possible outcomes per measurement. Then, in a typical Bell experiment, Alice and Bob each receive a part of a jointly prepared physical system, and then each of the parties independently choose one measurement to be performed on their individual system. Denoting Alice's measurement choice by $x$, and Bob's by $y$, we might describe such a Bell experiment by probability distributions $p_{xy}$ on the possible outcomes of $x$ and $y$. Repeating the experiment many times, Alice and Bob are able to estimate such probabilities.

Notice that a Bell scenario nicely falls into the description of a contextuality scenario. In this case, the contexts are given by a pair $xy$ composed of one measurement of Alice and one measurement of Bob. That is, the compatibility relations are granted by the spatial separation between the observers. Then, the no-disturbance conditions \eref{eq:no-disturbance} lead to the so-called \textit{no-signaling} conditions, which state that the probability distribution describing Alice's (Bob's) local experiment does not depend on Bob's (Alice's) measurement choice. Mathematically, the equation \eref{eq:no-disturbance} for this scenario becomes:
\numparts
\begin{align}
    \sum_b p_{x,y}(a,b) &= \sum_{b'} p_{x,y'}(a,b'), \\
    \sum_a p_{x,y}(a,b) &= \sum_{a'} p_{x',y}(a',b).
\end{align}   
\label{eq:no-signaling}
\endnumparts

The noncontexutal behaviors of a Bell scenario are named \textit{local} behaviors, and they satisfy:
\begin{equation}
    p_{xy}(a, b) = \sum_\lambda q(\lambda)p_{x}^\lambda(a) p_{y}^\lambda (b),
    \label{eq:usual-locality}
\end{equation}
where $a$ ($b$) denotes the possible outcomes of the measurement $x$ ($y$), while $q(\lambda) \geq 0$ and $\sum_{\lambda} q(\lambda) = 1$, here and in every subsequent use of this notation. 

On the other hand, in the so-called a \textit{generalized} Bell scenario, instead of only performing one measurement at each round of the experiment, Alice and/or Bob are allowed to jointly perform compatible measurements on their local systems. 
For simplicity, we restrict the discussion to scenarios where Alice still performs one measurement at each round, and only Bob has compatible measurements. 
Thus, in a generalized Bell scenario of this kind, a behavior is given by probability distributions $p_{x, C_B}$, where $x$ denotes Alice's single measurement and $C_B$ denotes a context of Bob's experiment, \textit{i.e.}, a set of compatible measurements he is able to jointly perform.

A generalized Bell scenario also can be seen as a contextuality scenario according to our previous discussion. In this case, the no-disturbance conditions represent both no-signaling conditions and no-disturbance conditions on Bob's local experiment. Mathematically, it means that, for every contexts $C_B,C'_B$ of Bob with no null intersection and every measurements $x,x'$ of Alice,
\numparts
\begin{align}
     p_{x,C_B\vert C_B \cap C_B'} & = p_{x,C_B'\vert C_B \cap C_B'},\\ 
     \sum_a p_{x,C_B}(a,\boldsymbol{b}) & =  \sum_{a'} p_{x',C_B}(a',\boldsymbol{b}),
\end{align}
where $\boldsymbol{b}$ denotes the outcome of Bob's context $C_B$, while, for every  $C_B,C'_B$ of Bob with null intersection,
\begin{equation}
    \sum_{\boldsymbol{b}} p_{x,C_B}(a,\boldsymbol{b}) = \sum_{\boldsymbol{b}'} p_{x,C'_B}(a,\boldsymbol{b}'),
\end{equation}
for every measurement $x$ and outcome $a$ for Alice. 
Thus, we refer to the set of acceptable behaviors in generalized Bell scenarios as $\mathcal{NSND}$ (nonsignaling and nondisturbing) polytope.
\endnumparts

For the sake of nomenclature, when dealing with the whole scenario as a single contextuality setup, we will refer to noncontextual behaviors as \textit{classical} behaviors. Thus, classical behaviors are convex combinations of nonsignaling and nondisturbing deterministic ones. Or, equivalently, a behavior $p_{x, C_B}$ is said to be classical if it admits a decomposition of the form
\begin{equation}
    p_{x, C_B} (a, \boldsymbol{b}) = \sum_\lambda q(\lambda)p_{x}^\lambda(a) \prod_{M \in C_B}p_{M}^\lambda (b_M),
    \label{eq:classical}
\end{equation}
where $b_M$ denotes the corresponding outcome of measurement $M \in C_B$.

From now on, in the generalized Bell scenarios here considered the term `contextuality' will be used in reference to local contextuality on Bob's marginal experiment. That is, a behavior of the \textit{generalized} scenario will be said to be noncontextual if \textit{Bob's marginal} behavior is noncontextual.

We should also be careful with the notions of locality for the generalized Bell scenarios. To begin with, notice that the definition \eref{eq:usual-locality} can also be applied to such scenarios. To do so, however, we cannot consider the full behavior $p_{x, C_B}$ at once. Rather, we should marginalize it to consider only one measurement $y$ of Bob at a time. We will refer to this as the \textit{usual} notion of locality.

Nevertheless, since Bob is typically performing more than one measurement at each round of the experiment, there might be correlations between Alice and a whole context of Bob. 
In this sense, in Ref. \cite{Temistocles2019} the authors propose a \textit{generalized} definition of locality, given by:
\begin{equation}
    p_{x, C_B} (a, \boldsymbol{b}) = \sum_\lambda q(\lambda)p_{x}^\lambda(a) p_{C_B}^\lambda (\boldsymbol{b}),
    \label{eq:generalized-local}
\end{equation}
where $\boldsymbol{b}$ denotes a possible (joint) outcome for the context $C_B$. 
Here, we ask for $p_{C_B}^\lambda$ to be only a \textit{nondisturbing} behavior of Bob's setup, and this is the key difference when this definition is compared with equation \eref{eq:classical}, where the response functions of Bob are also required to be noncontextual \cite{Mazzari23}. 
Furthermore, notice that if a behavior is local according to the generalized definition \eref{eq:generalized-local}, it is also local in the usual sense, given by equation \eref{eq:usual-locality}, considering any of the appropriate marginalizations. 

Let us also stress that the generalized definition of locality is better suited to the generalized scenarios than the usual one, as it takes the local compatibility structure of Bob into account. This will be the main notion of locality adopted in this work, and it is to be understood that whenever we mention `locality' in a generalized Bell scenario, we are referring to the generalized definition. Moreover, the inequalities associated with this definition of locality will be referred to as \textit{generalized Bell} inequalities.

\subsection{Relations between nonlocality and contextuality in generalized Bell scenarios} 
\begin{figure}
    \centering
    \includegraphics[width=0.9\textwidth]{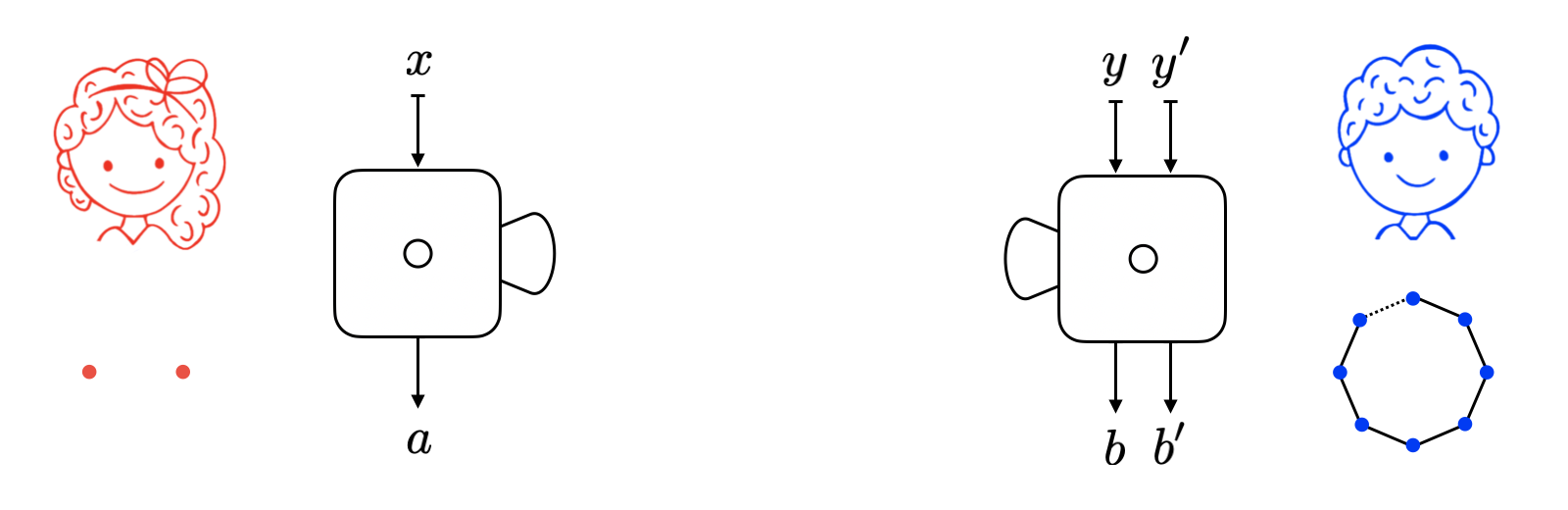}
    \caption{This figure represents a scenario where Alice is able to perform two incompatible measurements 
 -- represented by the two red dots -- and Bob is able to perform $n$ measurements which are pairwise compatible according to an $n$-cycle -- represented by the blue dots linked in a polygon-like compatibility graph. The label $x$ represents Alice's measurement choice, with respective outcome represented by $a$; labels $y$ and $y'$ represent a pair of compatible measurements of Bob, with respective outcomes $b$ and $b'$.}
    \label{fig:scenario1}
\end{figure}

Let us now move on to the discussion of some of the recent results relating nonlocality and contextuality in generalized Bell scenarios. Most of them have to do with a particular kind of the scenarios we discussed in the previous section, where the contextuality setup of Bob is an $n$-cycle \cite{Araujo2013}. That is, Bob is able to perform $n$ possible measurements $\{B_0, B_1, ..., B_{n-1}\}$, with contexts $\{B_i, B_{i+1}\}$ (sum modulo $n$), as depicted in Fig.~\ref{fig:scenario1}. All the measurements $B_i$, as well as Alice's measurements, that from now on will be denotet by $A_x$, with $x = 0, 1$, are dichotomic, with outcomes labeled by $+1$ and $-1$. 

One of the reasons why an $n$-cycle contextuality scenario is interesting is the fact that the nondisturbing polytope has been completely characterized in terms of its vertices, and the noncontextual polytope has been completely characterized in terms of its inequalities. Moreover, the $5$-cycle corresponds to the KCBS scenario \cite{kcbs}, which is the simplest scenario for which quantum theory exhibits contextuality. These generalized scenarios, where Alice has two incompatible measurements and Bob has an $n$-cycle setup, are also the focus of this work.

In Ref.~\cite{Kurzynski2014}, the authors consider precisely the scenario in which Bob has a KCBS setup. Then, they analyze violations of a KCBS inequality on Bob's experiment,
\begin{equation}
        \beta_{KCBS} = \langle B_0 B_1 \rangle + \langle B_1 B_2 \rangle + \langle B_2 B_3 \rangle  + \langle B_3 B_4 \rangle - \langle B_4 B_0 \rangle \leq 3,
    \label{eq:KCBS}
\end{equation}
together with a CHSH inequality of the form:
\begin{equation}
    \alpha_{CHSH} = \langle A_0 B_i \rangle + \langle A_0 B_j \rangle + \langle A_1 B_i \rangle - \langle A_1 B_j \rangle \leq 2,
    \label{eq:usualCHSH}
\end{equation}
where $B_i$ and $B_j$ are incompatible measurements of Bob. Notice that this CHSH inequality is associated with the usual notion of locality.  

Using the techniques of Ref.~\cite{Ramanathan2012}, it is possible to prove that for every nondisturbing behavior of the generalized scenario, the following relation holds:
\begin{equation}
    \alpha_{CHSH} + \beta_{KCBS} \leq 5.
    \label{eq:strict_monogamy}
\end{equation}
In other words, inequalities \eref{eq:KCBS} and \eref{eq:usualCHSH} cannot be simultaneously violated in such an experiment \cite{Kurzynski2014}.

Furthermore, notice that all the Bell inequalities associated with the usual notion of locality in such a scenario are CHSH inequalities of the form \eref{eq:usualCHSH}, up to relabelings \cite{Pironio2005}. Also, all the noncontextuality inequalities for Bob's experiment are known, and are also similar to \eref{eq:KCBS} up to relabelings.

The argument leading to equation \eref{eq:strict_monogamy} can then be applied to any of those CHSH inequalities together with any of the noncontextuality inequalities of the scenario. Therefore, if Alice and Bob share nonlocal correlations (according to the usual sense of locality), then Bob's local experiment cannot exhibit contextuality. Conversely, if Bob's experiment exhibits contextuality, then the correlations with Alice are local. That is, there is a fundamental monogamy relation between contextuality and the usual notion of locality in this scenario. The same argument was later extended to scenarios where Bob can perform joint measurements according to an arbitrary $n$-cycle \cite{jia2016}.

However, as we discussed in the previous section, the usual definition of locality, the one considered to prove this monogamy, is not the most appropriate notion of locality for generalized Bell scenarios. Thus, it is intersting to investigate whether this monogamy still holds when considering the generalized definition of locality proposed in Ref.~\cite{Temistocles2019}.

In Ref.~\cite{xue2023}, the authors present a negative answer to this question. They show a family of quantum states exhibiting both local contextuality on Bob's experiment and nonlocality in the generalized sense, by violating  the generalized Bell CHSH inequality
\begin{equation}
    \alpha_{CHSH} = \langle A_0 B_iB_{i+1} \rangle + \langle A_0 B_j \rangle + \langle A_1 B_iB_{i+1} \rangle - \langle A_1 B_j \rangle \leq 2,
    \label{eq:genCHSH}
\end{equation}
together with KCBS inequality \eref{eq:KCBS}.
The conclusion, then, is that there is not a strict monogamy relation between contextuality and nonlocality when the adequate definitions are considered.

Building up on this result, the main question we investigate in this manuscript is whether one can jointly observe arbitrary violations of nonlocality and contextuality in the scenarios studied therein, or if, in spite of the strict monogamy being disproved, there would still exist a trade-off relation between these two features.

\section{The trade-off in the simplest scenarios}\label{sec:trade-off}

Let us start by only considering the simplest generalized scenarios that have been explored in literature so far, i.e., the bipartite scenarios where Alice has only two dichotomic measurements, and Bob has a $n$-cycle setup, with sufficiently small $n$. A discussion involving more general scenarios is left to Section \ref{sec:more_scenarios}.

To investigate whether arbitrary violations of generalized Bell and noncontextuality inequalities can be observed in such scenarios, it is useful to adopt a notion of normalization for inequalities. For our purposes, it will be convenient to consider that a generalized-Bell (noncontextuality) inequality $S$ is normalized when $S \leq 0$ for all local (noncontextual) behaviors, and the maximum achieved by $\mathcal{NSND}$ behaviors is $S = 1$. Notice that every inequality can be normalized by appropriate rescaling and addition of an offset.

\begin{result}\label{result: 345tradeoff}
    Consider a generalized bipartite Bell scenario where Alice has two incompatible measurements and Bob has an $n$-cycle contextuality setup, with $n = 3$, $4$ or $5$. For any normalized generalized-Bell inequality $S_{GBell}$ and any normalized noncontextuality inequality $S_{NC}$ of Bob's local experiment, every behavior $\boldsymbol{p} \in \mathcal{NSND}$ satisfies
    \begin{equation}\label{eq: trade-off_inequalities}
        S_{GBell}(\boldsymbol{p}) + S_{NC}(\boldsymbol{p}) \leq 1.
    \end{equation}
\end{result}

The proof of this result goes through an evaluation of all the vertices of the scenarios' $\mathcal{NSND}$ polytopes, and is discussed in details in  \ref{app:proof_tradeoff_ineqs}. The restriction to $n \leq 5$ is solely due to computational limitations.

This result shows that in the simplest generalized Bell scenarios, the no-signaling and no-disturbance conditions bound the joint amount of nonlocality and contextuality a behavior might have. This bound is not as strict as firstly proposed in Ref.~\cite{Kurzynski2014}, but it still imposes a fundamental trade-off relation between these two features. 

In particular, it is interesting to notice that the maximum value of $S_{GBell} + S_{NC}$ allowed by no-signaling and no-disturbance may be achieved having only one of nonlocality or contextuality. Moreover, according to expression \eref{eq: trade-off_inequalities}, a maximal violation of a generalized-Bell inequality implies that Bob's marginal is noncontextual, and a maximal violation of a noncontextual inequality in Bob's local experiment also implies that the correlations between Alice and Bob are local. Thus, in the considered scenarios, a maximum amount of nonlocality implies noncontextuality, and vice-versa.

Another interesting consequence of such a trade-off relation is the fact that it forbids the existence of an inequality which is only violated by behaviors which contain both nonlocality and local contextuality.

\begin{result}\label{result: no_joint_witness}
   In a given generalized Bell scenario, the following two statements are equivalent: i) Expression \eref{eq: trade-off_inequalities} holds for every pair of generalized-Bell and noncontextuality inequalities; ii) There is no joint witness of nonlocality and (local) contextuality.
\end{result}

To see this, notice that if this witness were to exist, its maximum value in the $\mathcal{NSND}$ polytope would be achieved in a nonlocal and contextual extremal point. However, if there would be such an extremal point, it would maximally violate a generalized-Bell inequality, and, simultaneously, it would violate some noncontextuality inequality, contradicting \eref{eq: trade-off_inequalities}. The converse can be seen by noticing that if some behavior does not respect inequality \eref{eq: trade-off_inequalities} for some generalized-Bell and some noncontextuality inequality, then \eref{eq: trade-off_inequalities} is itself a witness of both non-locality and contextuality. 

This result can be related to the phenomenon of `nonlocality revealed by local contextuality' discussed in Refs.~\cite{cabello2010, Saha2016}, for which the existence of an inequality that can only be violated by behaviors containing nonlocality and contextuality is required. Thus, the above result states that such a phenomenon cannot exist in scenarios satisfying the trade-off \eref{eq: trade-off_inequalities}.

From a more applied point of view, since nonlocality and contextuality are both resources for information processing tasks, a joint observation of them in the generalized scenarios hinted that these scenarios could offer novel possibilities for such applications, using both resources together \cite{xue2023}. However, the existence of this trade-off relation between them indicates that we may not obtain advantages coming from considerable amounts of nonlocality and contextuality concomitantly. In addition, considering the comment of the last paragraph, one might ask whether the joint presence of nonlocality and contextuality brings any advantage at all. 

On the other hand, there might be possibilities for practical applications of the generalized scenarios based on this trade-off. For instance, if Bob certifies a given amount of contextuality he locally posess, then he can bound the amount of nonlocal correlations he has with other observers. This is a promising research line that requires deeper investigation.

\section{Analysing the trade-off with quantifiers}\label{sec:trade-off-quantifiers}

Inequality \eref{eq: trade-off_inequalities}, discussed in the last section, expresses a fundamental trade-off relation between nonlocality and contextuality in the simplest generalized Bell scenarios, as it does not involve a particular choice of generalized-Bell or noncontextuality inequalities. In fact, this relation can be equivalently described in terms of \textit{quantifiers} of nonlocality and contextuality. We begin this section by defining quantifiers of nonlocality and contextuality suitable for generalized Bell scenarios, and rewriting the trade-off relation in terms of them. By doing so, we will get significant insights both on the relation itself and on the actual meaning of nonclassical correlations in generalized Bell scenarios.

Naturally extending the definitions in Refs.~\cite{elitzur92, Abramsky2017} to generalized Bell scenarios, the \textit{nonlocal fraction} of a behavior $\boldsymbol{p}$ belonging to the $\mathcal{NSND}$ polytope of a given scenario, denoted by $NLF(\boldsymbol{p})$, is defined by:
\begin{equation}\label{def:nl_fraction}
    NLF(\boldsymbol{p}) = \min \{\lambda \in [0, 1] \vert \boldsymbol{p} = (1 - \lambda) \boldsymbol{p}_{L} + \lambda \boldsymbol{p}_{NSND}\},
\end{equation}
where the minimization is taken over all local behaviors $\boldsymbol{p}_{L}$ and all behaviors $\boldsymbol{p}_{NSND}$ belonging to the $\mathcal{NSND}$ polytope. 
In a similar way, for the generalized scenarios here considered, where only Bob has compatible measurements, the \textit{contextual fraction} of a behavior $\boldsymbol{p}$, denoted by $CF(\boldsymbol{p})$, is defined by: 
\begin{equation}\label{def:contextual_fraction}
    CF(\boldsymbol{p}) = \min\{\lambda \in [0, 1] \vert \boldsymbol{p}_B = (1 - \lambda) \boldsymbol{p}_{B_{NC}} + \lambda \boldsymbol{p}_{B_{ND}}\},
\end{equation}
where the minimization is taken over all noncontextual behaviors $\boldsymbol{p}_{B_{NC}}$ and all nondisturbing behaviors $\boldsymbol{p}_{B_{ND}}$ of Bob's local experiment \footnote{Recall that in the scenarios we study in this work, the contextuality properties of a behavior are associated to Bob's local experiment. That is the reason why define the contextual fraction in this fashion.}.

It is also useful to define a quantifier associated with a global notion of contextuality for the generalized Bell scenarios. As we mentioned in Subsection \ref{sec:bell_scenarios}, in order to avoid confusions with the notion of local contextuality on Bob's local experiment, we refer to the `contextuality in the whole scenario' as \textit{nonclassicality}.

Thus, the \textit{nonclassical fraction} of a behavior $\boldsymbol{p}$, denoted by $NClF(\boldsymbol{p})$, is defined by:
\begin{equation}
    NClF(\boldsymbol{p}) = min\{\lambda \in [0, 1] \vert \boldsymbol{p} = (1 - \lambda) \boldsymbol{p}_{Cl} + \lambda \boldsymbol{p}_{NSND}\},
\end{equation}
where the minimization runs over all classical behaviors $\boldsymbol{p}_{Cl}$ (see equation \eref{eq:classical}) and all behaviors $\boldsymbol{p}_{NSND}$ belonging to the $\mathcal{NSND}$ polytope.

Considering these definitions, the trade-off given by \eref{eq: trade-off_inequalities} can be expressed in terms of quantifiers according to the following result, the proof of which can be found in  \ref{app:proof_tradeoff_nonclassical}.

\begin{result}\label{result:tradeoff_nonclassical}
     In a given generalized Bell scenario, the following two statements are equivalent: i) Expression \eref{eq: trade-off_inequalities} holds for every pair of generalized-Bell and noncontextuality inequalities; ii) The quantifiers nonlocal fraction ($NLF$), contextuality fraction ($CF$), and nonclassical fraction ($NClF$) satisfy the trade-off relation:
\begin{equation}\label{eq:trade-off_nonclassical}
        NLF(\boldsymbol{p}) + CF(\boldsymbol{p}) \leq NClF(\boldsymbol{p}).
    \end{equation}
\end{result}     

This suggests that nonlocality and contextuality can be seen as two distinct manifestations of a broader notion of nonclassicality for the generalized scenarios. Thus, the trade-off between them follows from the fact that the no-signaling and no-disturbance conditions limit the amount of nonclassicality allowed in such scenarios.

However, a particularly intriguing aspect of expression \eref{eq:trade-off_nonclassical} is the fact that it may not be an equality, \textit{i.e.}, it indicates that the sum of the nonlocal and contextual fractions may be strictly smaller than the nonclassical fraction of a given behavior. 
In other words, it suggests that \emph{nonlocality and contextuality are not the only kinds of nonclassical correlations in generalized Bell scenarios}. 
This aspect is discussed in more details in  \ref{app:nonclassicality_example}, where we provide an explicit example of a behavior for which the strict inequality in equation \eref{eq:trade-off_nonclassical} holds. In particular, this behavior is local, noncontextual, but still exhibits some form of nonclassicality, which has a nice operational interpretation.

We should note that expression \eref{eq:trade-off_nonclassical} resembles a similar inequality for quantifiers of entanglement and contextuality of quantum states, obtained in Ref.~\cite{Camalet2017}. However, notice that in this work our approach is completely theory-independent, and inequality \eref{eq:trade-off_nonclassical} is valid for all nonsignalling and nondisturbing behaviors in the considered scenarios. Meanwhile, in Ref.~\cite{Camalet2017} the author works in the scope of quantum theory, and the result is related to concepts only well defined therein, like the entanglement of a given quantum state. 
Also, it is worth noting that entanglement and nonlocality are related but not equivalent concepts, see, for example, Ref.~\cite{Methot2008}.

\begin{figure}[H]
    \centering
    \includegraphics[width=\textwidth]{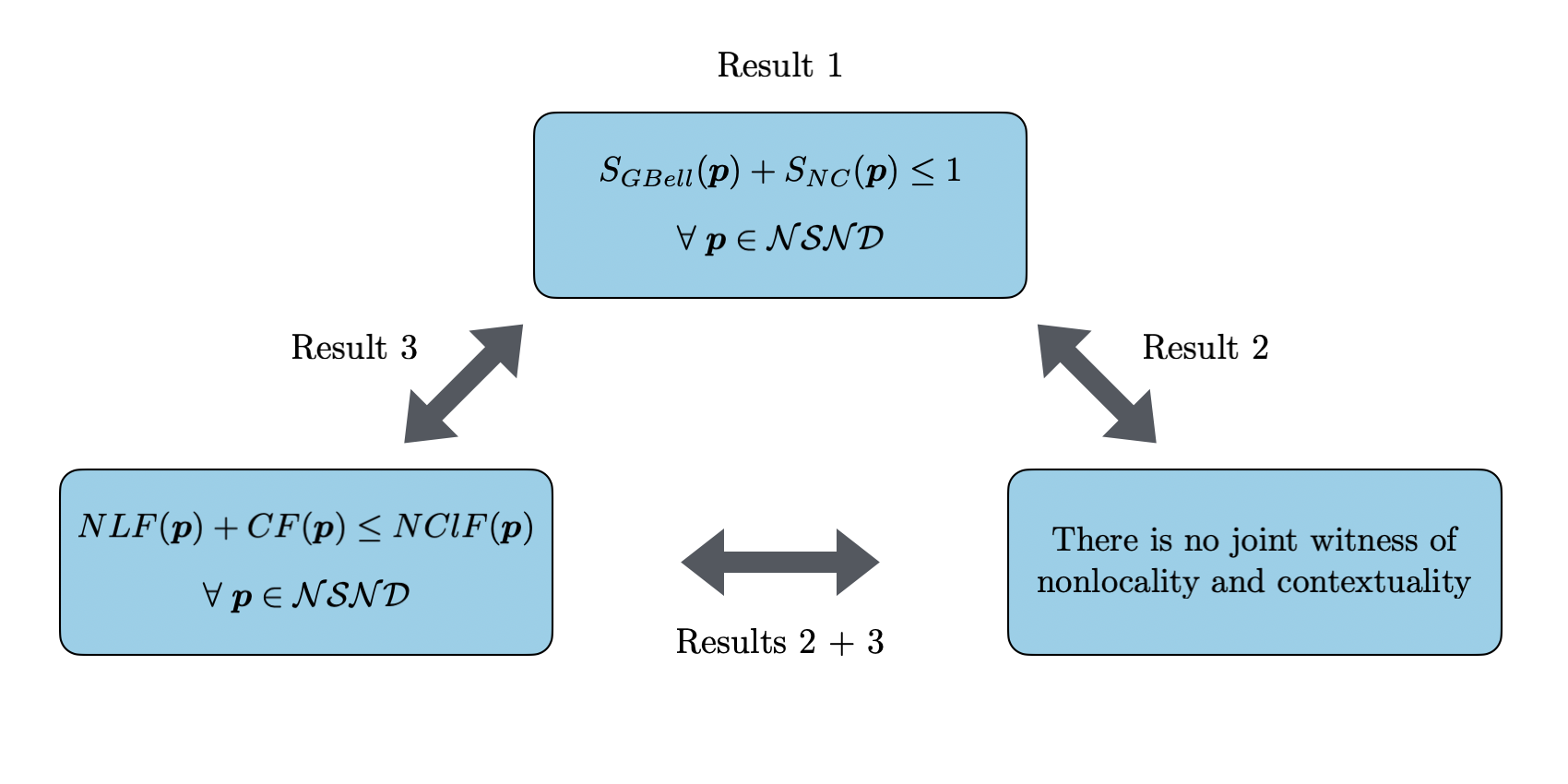}
    \caption{A diagram summarizing the relation between the results presented in previous sections, valid for scenarios where Alice measures two binary measurements and Bob measures an $n$-cycle scenario for $n=3,4,5$. Result \ref{result: 345tradeoff} refers to the trade-off relation with normalized inequalities; result \ref{result: no_joint_witness} shows that there is no convex witness of nonlocality and contextuality; and result \ref{result:tradeoff_nonclassical} refers to the trade-off expressed in terms of quantifiers.}
    \label{fig:digram}
\end{figure}

Relations between the results are depicted in Fig.~\ref{fig:digram}.

\section{Other scenarios}\label{sec:more_scenarios}

In the previous sections, our discussion has been primarily focused on exploring the trade-off relation described in Result \ref{result: 345tradeoff}, which is valid for the simplest generalized Bell scenarios. Now, we aim at studying the validity of this result in a wider class of bipartite scenarios.

\subsection{The n-cycle scenarios}

To begin this study, let us consider scenarios where Bob has an arbitrary $n$-cycle, which are natural extensions of the scenarios for which Result \ref{result: 345tradeoff} holds.

\begin{conjecture}\label{conj:all_ncycles}
    Consider a bipartite generalized Bell scenario where Bob has an $n$-cycle contextuality setup. For any normalized generalized-Bell inequality $S_{GBell}$ and any normalized noncontextuality inequality $S_{NC}$ of Bob's local experiment, every $\mathcal{NSND}$ behavior satisfies expression \eref{eq: trade-off_inequalities}.
\end{conjecture}
This conjecture is mainly supported by two reasons. First, we can prove that in any of these scenarios maximal contextuality implies locality, according to the following result.

\begin{result}\label{result: max_contextual_implies_local}
    Consider a bipartite generalized Bell scenario where Bob has an $n$-cycle contextuality setup. If a $\mathcal{NSND}$ behavior maximally violates a noncontextuality inequality of Bob's marginal experiment, then it must be local.
\end{result}
The proof of this result can be found in \ref{app:proof_maximal_contextual_implies_local}. It is mostly based on the fact that we know all the noncontextuality inequalities for the $n$-cycle contextuality scenarios \cite{Araujo2013}, and they are such that there is only one vertex achieving a maximal violation of each of them.

On the other hand, the second reason why we believe in Conjecture \ref{conj:all_ncycles} has to do with some of the generalized-Bell inequalities showing up on those scenarios. Unlike the noncontextuality inequalities of $n$-cycle contextuality scenarios, there is not a clear characterization of the Bell inequalities in these generalized scenarios. Even so, at least two of them are straightforward extensions of well-known and studied Bell inequalities, namely, the CHSH inequality \cite{CHSH, Temistocles2019, xue2023} (see equation \eref{eq:genCHSH}) and the so-called chained inequality \cite{scaranibell, chained_ineqs}. 
In  \ref{app:example1} and \ref{app:example2}, we properly define these inequalities, and we prove that a maximal violation of them in the scenarios where Bob has an $n$-cycle setup implies that his marginal is noncontextual.

These results show that the $n$-cycle contextuality scenarios are deeply connected with the trade-off relation between nonlocality and contextuality observed in generalized Bell scenarios. The main feature of these contextuality scenarios that seems to be responsible for such a phenomenon is the fact that the only maximally contextual behaviors are vertices of its associated $\mathcal{ND}$ polytope, which, as discussed in  \ref{app:proof_maximal_contextual_implies_local}, is crucial for the proof of Result \ref{result: max_contextual_implies_local}.

In contrast, let us now discuss an example of a generalized Bell scenario where Bob's local compatibility structure is related to another well-studied proof of contextuality, the so-called \textit{Peres-Mermin square}  \cite{peres90, mermin1990, Budroni2021}.

\subsection{The Peres-Mermin square scenarios}

\begin{figure}
    \centering
    \includegraphics[width=0.6\textwidth]{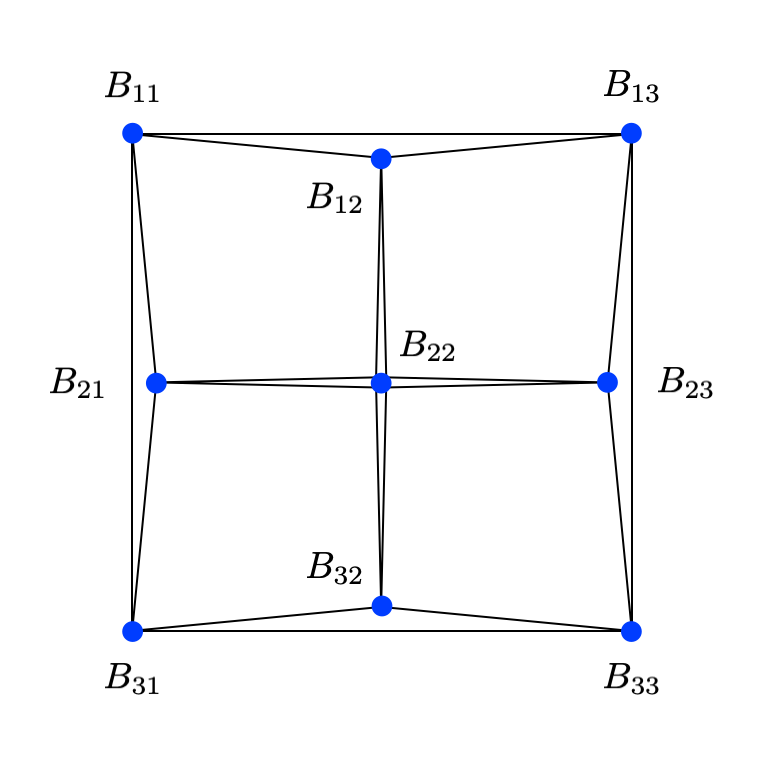}
    \caption{Compatibility graph for the Peres-Mermin square contextuality scenario. There are nine measurements (represented by the blue dots), eighteen contexts of length two (represented by edges connecting the dots) and six contexts of length three (represented by triangles connecting three dots).}
    \label{fig:peres}
\end{figure}

The Peres-Mermin square is a proof of contextuality which takes place in a scenario with nine dichotomic measurements with compatibility structures illustrated by figure \ref{fig:peres}. It has been extensively studied in literature for various reasons, for instance, it can exhibit state-independent quantum contextuality \cite{cabello2008}, and it is the basis for the simplest Bell scenario where quantum theory exhibits full nonlocality \cite{brassard2005, aravind2002, aravind2002b}.

\begin{figure}[H]
    \centering
    \includegraphics[width=0.9\textwidth]{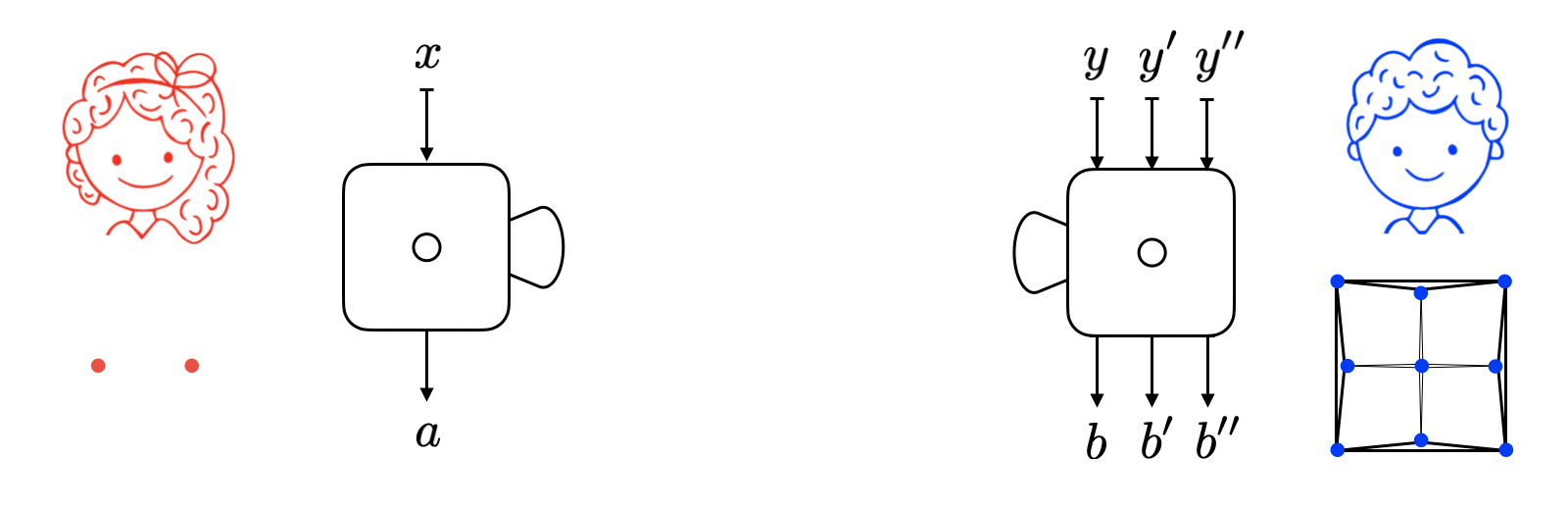}
    \caption{This figure represents a scenario where Alice is able to perform two incompatible measurements 
 -- represented by the two red dots -- and Bob is able to perform $9$ measurements which are triple-wise compatible according to a Peres-Mermin square -- represented by the blue dots linked according to the compatibility graph of the Peres-Mermin square (Fig.~\ref{fig:peres}). The label $x$ represents Alice's measurement choice, with respective outcome represented by $a$; labels $y$, $y'$ and $y''$ represent a triple of compatible measurements of Bob, with respective outcomes $b$, $b'$, and $b''$.}
    \label{fig:scenario2}
\end{figure}

In the context of bipartite generalized Bell scenarios, if Bob has a Peres-Mermin square and Alice has at least two incompatible measurements (such as represented in Fig.~\ref{fig:scenario2}), there is no trade-off whatsoever between nonlocality and contextuality. That is, there are $\mathcal{NSND}$ behaviors which maximally violate a noncontextuality inequality and a generalized-Bell inequality simultaneously. In  \ref{app:peres-mermin}, we explicitly show an example of such a behavior.

In the first place, the existence of these behaviors shows that the trade-off between nonlocality and contextuality is not a universal phenomenon that holds in every generalized Bell scenario. Surprisingly, this means that expression \eref{eq:trade-off_nonclassical} does not always hold. That is, there exists a behavior for which the sum of its nonlocal and contextual fractions overcomes its nonclassical fraction, which seems to indicate that accounting for nonlocality and contextuality separately leads to a redundancy in the quantification of nonclassical correlations. This points out possibilities of correlations which are only nonclassical if nonlocality and contextuality are jointly present, an aspect that we leave for future investigations.

Moreover, the fact that the trade-off described by \ref{result: 345tradeoff} does not hold in every generalized Bell scenario further highlights the connection between this pheonomenon and the $n$-cycle contextuality scenarios, and it raises the question of which features of the $n$-cycle scenarios, which cannot be present in the cases when Bob has a Peres-Mermin square, lead to the trade-off. As mentioned in the last subsection, an apparently good guess for such a feature is the fact that in the $n$-cycle contextuality scenarios the only behaviors which maximally violate noncontextuality inequalities are vertices of the associated $\mathcal{ND}$ polytope. This turns out not to be the case in the Peres-Mermin contextuality scenario\footnote{This can be seen from the fact that fixing the measurements in the Peres-Mermin square, this scenario exhibits a noncontextuality inequality that can be maximally violated by any quantum state \cite{cabello2008, Budroni2021}. Since not all quantum states lead to the same behavior, it follows that there exists more than one behavior which is able to maximally violate such an inequality.}, further strengthening our belief in the crucial role performed by this fact for the trade-off to exist. A deeper investigation of this question, and a complete characterization of the generalized Bell scenarios where the trade-off holds is left for future research.

\section{Final Remarks}\label{sec:conclusion}

In this work, we further analysed the relations between nonlocality and contextuality in generalized Bell scenarios. Recently, a joint observation of nonlocality and contextuality in a scenario where they were thought to be monogamous has been reported \cite{Kurzynski2014, xue2023}. Here, we investigate whether this joint observation can be arbitrary, or if there still exist a certain trade-off relation between nonlocality and contextuality in the considered generalized scenarios.

At least for the simplest of them, we proved that the latter is indeed the case, and thus one cannot expect to concomitantly observe arbitrary amounts of nonlocality and contextuality. By rewriting the trade-off relation in terms of quantifiers, we discussed how nonlocality and contextuality can be seen as two distinct manifestations of nonclassical correlations, and the trade-off relation then follows from a limitation of the amount of such correlations imposed by the no-signaling and no-disturbance conditions.

Although this provides some intuition on the reason why the trade-off exists, an interesting problem that this work leaves open is the one of understanding the relations between nonlocality and contextuality from a more physical perspective. The arguments we have shown take into account mainly the geometry of the scenarios. 
However, studying the relations between these features from a more physical perspective would guide us towards a more foundational understanding of them. The underlying reason for the trade-off, for instance, might be related to interplays between strongly nonlocal correlations and local randomness \cite{scaranibell}.

An attempt to investigate this question may certainly benefit from a characterization of which generalized Bell scenarios demands for trade-off relations. 
In this work we also give some first steps in this direction, by providing strong evidence that the trade-off exists whenever Bob has a $n$-cycle contextuality setup, and also by showing that this is not the case if he has a Peres-Mermin square. However, for a complete charaterization to be obtained further research is required.

Our results can be connected to an existing discussion in the literature related to the notion of `nonlocality revealed by local contextuality' \cite{cabello2010, liu2016, Saha2016}. The trade-off between nonlocality and contextuality exhibited in Result \ref{result: 345tradeoff} is a radically different phenomenon from the revelation of nonlocality by local contextuality. In the former, it seems like nonlocality and contextuality are two distinct features coming from a common origin, and the trade-off follows from a limitation on such an origin. In the latter, it seems like nonlocality and contextuality are related in such a way that one cannot exist without the other. It is important to stress, however, that the notion of nonlocality revealed by local contextuality has never been studied in terms of the generalized definition of locality. This would be essential for us to better understand how the relation between nonlocality and contextuality differs from one scenario to another.

Finally, from a more practical perspective, the trade-off between nonlocality and contextuality implies mainly two considerations. On the one hand, it shows that in some scenarios one cannot use arbitrary amounts of nonlocality and contextuality jointly, thus imposing a fundamental limit on practical applications based on both of them. On the other hand, a trade-off can also inspire other possibilities for applications, such as bounding the amount of nonlocal correlations from an estimated amount of local contextuality. Since nonlocality and contextuality are resources for many information-processing tasks, combining their use in such scenarios is a prominent research endeavor, which must be thoroughly explored in the near future.

\section*{Acknowledgements}
The authors thank Carlos Vieira and Pedro Lauand for interesting discussions. 
This work was supported by the S\~{a}o Paulo Research Foundation FAPESP (grants nos.\ 2018/07258-7, 2021/10548-0, 2023/04197-5, 2023/12979-3 and 2023/04053-3), the Brazilian national agency CNPq (grants nos.\ 310269/2019-9, 311314/2023-6, 316657/2023-9) and it is part of the Brazilian National Institute for Science and Technology in Quantum Information. PK is supported by the Polish National Science Centre (NCN) under the Maestro Grant no. DEC-2019/34/A/ST2/00081.

\section*{References}
\bibliography{final_version}
\bibliographystyle{unsrt}
\appendix
\section{Proof of Result \ref{result: 345tradeoff}}\label{app:proof_tradeoff_ineqs}
To start this proof, notice that in a given generalized Bell scenario, expression \eref{eq: trade-off_inequalities} holds if, and only if, none of the vertices of the corresponding $\mathcal{NSND}$ polytope is simultaneously nonlocal and contextual. Then, to obtain the vertices of the considered scenarios we use PANDA \cite{PANDA} to convert the polytope's half-space description (which can be obtained from positivity and no-signaling conditions) to its vertex description. The code details and the list of vertices for the three scenarios $n = 3$, $4$ and $5$ are available at \cite{repo}.

Once we have a list of vertices, the local ones are easily identified since Alice's marginal is deterministic in all of them. All the others must be nonlocal. Now, for each nonlocal vertex, we check whether Bob's marginal is contextual. This can be done in two equivalent ways: one can either compute the values of all the noncontextuality inequalities' expressions, which are known for $n$-cycle scenarios \cite{Araujo2013}; or one can use linear programming techniques to try to write Bob's marginal as a convex combination of deterministic strategies \cite{Abramsky2017}. By doing so, we observe that \textit{all} nonlocal vertices of the $\mathcal{NSND}$ polytope for the considered scenarios are noncontextual \cite{repo}.

\section{Proof of Result \ref{result:tradeoff_nonclassical}}\label{app:proof_tradeoff_nonclassical}
For any given $\mathcal{NSND}$ behavior $\boldsymbol{p}$, we can write
\begin{equation}
    \boldsymbol{p} = NClF(\boldsymbol{p})\boldsymbol{p}_{NSND} + (1 - NClF(\boldsymbol{p}))\boldsymbol{p}_{Cl},
\end{equation}
where $\boldsymbol{p}_{NSND}$ is some $\mathcal{NSND}$ behavior, and $\boldsymbol{p}_{Cl}$ is a classical behavior.

The behavior $\boldsymbol{p}_{NSND}$, in particular, can be further decomposed into a convex combination of a local behavior and a \textit{strongly} nonlocal behavior (a behavior is said to be \textit{strongly} nonlocal if its nonlocal fraction is equal to one). 
Consequently, we can think of them as being a convex combination of nonlocal vertices of the $\mathcal{NSND}$ polytope. 
Thus, we may write
\begin{equation}
    \boldsymbol{p}  = \alpha \boldsymbol{p}_{SNL} + (NClF(\boldsymbol{p}) - \alpha)\boldsymbol{p}_L + (1 - NClF(\boldsymbol{p}))\boldsymbol{p}_{Cl},
\end{equation}
where $0 \leq \alpha \leq 1$. Moreover, from \eref{def:nl_fraction}, it follows that $\alpha \geq NLF(\boldsymbol{p})$.

Then, since \eref{eq: trade-off_inequalities} is satified for every $\mathcal{NSND}$ behavior, and every generalized-Bell and noncontextuality inequalities, it follows that Bob's marginal of every strongly nonlocal behavior must be noncontextual. Consequently, the equation above implies that Bob's marginal behavior satisfies
\begin{equation}
    \boldsymbol{p}_B = (NClF(\boldsymbol{p}) - \alpha)\boldsymbol{p}_{B_{ND}} + (1 + \alpha - NClF(\boldsymbol{p}))\boldsymbol{p}_{B_{NC}}.
\end{equation}
Therefore, from definition \eref{def:contextual_fraction} we conclude that $NClF(\boldsymbol{p}) - \alpha \geq CF(\boldsymbol{p})$, from which the result follows.

The converse follows from the fact that \eref{eq:trade-off_nonclassical} implies that every nonlocal vertex of the $\mathcal{NSND}$ polytope (which has unit nonlocal fraction, and unit nonclassical fraction) must be noncontextual. And, as discussed in  \ref{app:proof_tradeoff_ineqs}, this is equivalent to saying that every $\mathcal{NSND}$ behavior satisfies \eref{eq: trade-off_inequalities} for all inequalities.

\section{Are nonlocality and local contextuality the only forms of nonclassicality in generalized Bell scenarios?}\label{app:nonclassicality_example}

Consider a bipartite generalized Bell scenario where only Bob has compatible measurements, and let us assume that at each round of the experiment, after the measurements have been performed, Alice communicates her input and her output to Bob. 
Then, with this extra information Bob is able to improve his description of his local experiment, by constructing what we call \textit{conditional} behaviors. 
Starting from a nonsignaling and nondisturbing behavior of the whole experiment $p_{x, C_B}$, Bob's conditional behaviors are defined by $p^x_{C_B}(\boldsymbol{b}|a) = \frac{p_{x, C_B}(a, \boldsymbol{b})}{p_x(a)}$.

Now, assume that in such a scenario Alice and Bob share a classical behavior $p_{x, C_B}$, that is, a behavior that can be decomposed as in equation \eref{eq:classical}. 
Then, the conditional behaviors computed from it can be written as $p^x_{C_B} (\boldsymbol{b}|a) = \sum_\lambda \frac{q(\lambda)p_{x}^\lambda(a)}{p_x(a)} \prod_{M \in C_B}p_{M}^\lambda (b_M) $. 
However, for each $x$ and $a$, this is a decomposition of the form \eref{eq:noncontextuality}, meaning that Bob's conditional behaviors are noncontextual. In other words, if Alice and Bob share a classical behavior, all the conditional behaviors originated from it must be noncontextual.

With this in mind and now making use of quantum theory, let us now consider that Alice and Bob share the qubit-qutrit state
\begin{equation}\label{eq:state}
    \rho = \frac{1}{2}| 0 \rangle \langle 0 |_A \otimes | 0 \rangle \langle 0 |_B + \frac{1}{2} | 1 \rangle \langle 1 |_A \otimes | 2 \rangle \langle 2 |_B, 
\end{equation}
and Bob has a $5$-cycle structure, with measurements $B_j = 2|v_j \rangle \langle v_j|_B - \mathbf{I}_B$, where $|v_j \rangle_B = cos\theta |0\rangle_B + sin\theta cos(j4\pi/5)|1\rangle_B + sin\theta sin(j4\pi/5)|2\rangle_B$, with $cos^2 \theta = cos(\pi/5)/(1 + cos(\pi/5))$, for $j = 1, 2,3,4,5$. Moreover, suppose that Alice only performs one measurement, given by $A = \sigma_z$.

Since the state is separable, the behavior it originates has to be local (see Ref.~\cite{Temistocles2019}). Also, it is possible to verify that Bob's marginal behavior is noncontextual. To do so, we implement a linear program to calculate the noncontextual fraction of nondisturbing behaviors in a $5$-cycle scenario, according to Ref.~\cite{Abramsky2017}, and check that the noncontextual fraction of Bob's marginal behavior is equal to one \cite{repo}.

However, consider Bob's conditional behavior when Alice's measurement output is $+1$, that is, the conditional behavior $p_{C_B}(\boldsymbol{b}|a = +1)$. Since the state shared by Alice and Bob is the one given by \eref{eq:state}, this conditoinal behavior refers to the cases when Bob's marginal state is $|0\rangle_B$. Then, it is straightforward to verify that this state with the measurements $B_j$ above mentioned violates the noncontextuality inequality
\begin{equation}
        \langle B_0 B_1 \rangle + \langle B_1 B_2 \rangle + \langle B_2 B_3 \rangle + \langle B_3 B_4 \rangle + \langle B_4 B_0 \rangle \geq -3
\end{equation}
by $5 - 4\sqrt{5} \approx -3.94$ \cite{Budroni2021}. 
This implies that Bob's conditional behavior $p_{C_B}(\boldsymbol{b}|a = +1)$ is contextual, and, therefore, the original behavior in the generalized scenario is nonclassical.

Alternatively, another way to verify that this behavior is nonclassical is to show that it violates an inequality which is satisfied by all classical behaviors. For that matter, consider the following inequality
\begin{equation}\label{eq:classicality_ineq}
        3\mean{A} + \mean{(1+A)D} \geq -3,
\end{equation}
where $D = B_1B_2 +  B_2B_3 +  B_3B_4 +  B_4B_5 +  B_5B_1$. It is straightforward to check that every classical behavior satisfies it, by simply checking all the deterministic assignments of $\pm 1$ to the observables $A$ and $B_j$. For all such assignments, we have $ \mean{(1+A)D} = \mean{(1+A)}\mean{D}$ and $\mean{D} \geq -3$, from which the bound in \eref{eq:classicality_ineq} directly follows. On the other hand, the given quantum behavior violates \eref{eq:classicality_ineq} by $5 - 4\sqrt{5} \approx -3.94$, thus proving its nonclassicality.

\section{Proof of Result \ref{result: max_contextual_implies_local}}\label{app:proof_maximal_contextual_implies_local}

To prove Result \ref{result: max_contextual_implies_local}, we start by showing that, in an $n$-cycle contextuality scenario, the only behaviors which maximally violate noncontextuality inequalities are vertices of the associated nondisturbing polytope. To do so, let us recall that the contextual vertices of the nondisturbing polytope of such scenarios are given by:
\numparts
    \begin{equation}
        \langle B_i \rangle = 0
    \end{equation}
    \begin{equation}\label{eq:correlators_contextual_vertex_ND}
        \langle B_i B_{i+1} \rangle = \gamma_i,
    \end{equation}
\endnumparts
where $i = 1, 2, ..., n$ and $\gamma_i = \pm 1$, such that the number of $i$'s for which $\gamma_i = -1$ is odd \cite{Araujo2013}. For such vertices, the following lemma, which is proven in details in Ref.~\cite{MarcotulioDiss}, holds.

\begin{lemma}
    An equally weighted convex combination of any two contextual vertices of the $n$-cycle contextuality scenario is a noncontextual behavior.
\end{lemma}
{\em Proof:} Let $\boldsymbol{v}_1,\boldsymbol{v}_2$ be two contextual vertices of an $n$-cycle contextuality scenario and let $\boldsymbol{p}$ be the equally weighted convex combination of these vertices. Considering equations \eref{eq:correlators_contextual_vertex_ND}, all the correlators $\langle B_i B_{i+1} \rangle_{\boldsymbol{p}}$ must be either $-1$, $0$ or $1$. If $\langle B_i B_{i+1} \rangle_{\boldsymbol{v}_1}$ and $\langle B_i B_{i+1} \rangle_{\boldsymbol{v}_2}$ have both the same sign, we have either $\langle B_i B_{i+1} \rangle_{\boldsymbol{p}} = 1$ or $\langle B_i B_{i+1} \rangle_{\boldsymbol{p}} = -1$. On the other hand, if $\langle B_i B_{i+1} \rangle_{\boldsymbol{v}_1}$ and $\langle B_i B_{i+1} \rangle_{\boldsymbol{v}_2}$ have opposite signs, then $\langle B_i B_{i+1} \rangle_{\boldsymbol{p}} = 0$. Notice that the number $i$'s for which $\langle B_i B_{i+1} \rangle_{\boldsymbol{p}} = 0$ has to be even, since the number of $i$'s for which $\langle B_i B_{i+1} \rangle_{\boldsymbol{v}_1}$ and $\langle B_i B_{i+1} \rangle_{\boldsymbol{v}_2}$ have different signs is necessarily even. 

Now, if there is an even number of $\langle B_i B_{i+1} \rangle_{\boldsymbol{p}} = -1$, construct two noncontextual behaviors $\boldsymbol{p}_{NC_1}$ and $\boldsymbol{p}_{NC_2}$ according to the following rules: (a) if $\langle B_i B_{i+1} \rangle_{\boldsymbol{p}} = \pm 1$, then $\langle B_i B_{i+1} \rangle_{\boldsymbol{p}_{NC_1}} = \langle B_i B_{i+1} \rangle_{\boldsymbol{p}_{NC_2}} = \langle B_i B_{i+1} \rangle_{\boldsymbol{p}}$; and (b) if $\langle B_i B_{i+1} \rangle_{\boldsymbol{p}} = 0$, then $\langle B_i B_{i+1} \rangle_{\boldsymbol{p}_{NC_1}} = -\langle B_i B_{i+1} \rangle_{\boldsymbol{p}_{NC_2}} = 1$. Finally, one can check that if we assign $\langle B_i \rangle_{\boldsymbol{p}_{NC_1}} = \langle B_i \rangle_{\boldsymbol{p}_{NC_2}} = 0$, it follows that $\boldsymbol{p}_{NC_1}$ and $\boldsymbol{p}_{NC_2}$ are indeed valid noncontextual behaviors, since each of them has an even number of $\langle B_i B_{i+1} \rangle = -1$. 

Similarly, if there is an odd number of $\langle B_i B_{i+1} \rangle_{\boldsymbol{p}} = -1$, consider $\boldsymbol{p}_{NC_1}$ and $\boldsymbol{p}_{NC_2}$ constructed in the same way as described before. However, choose one $j$ such that $\langle B_j B_{j+1} \rangle_{\boldsymbol{p}} = 0$ and flip the signs of $\langle B_j B_{j+1} \rangle_{\boldsymbol{p}_{NC_1}}$ and $\langle B_j B_{j+1} \rangle_{\boldsymbol{p}_{NC_2}}$. Again, for the same reason as before $\boldsymbol{p}_{NC_1}$ and $\boldsymbol{p}_{NC_2}$ are valid noncontextual behaviors. 

In both of the cases described above, it holds that $\boldsymbol{p} = \frac{1}{2}\boldsymbol{p}_{NC_1} + \frac{1}{2}\boldsymbol{p}_{NC_2}$, which proves that $\boldsymbol{p}$ is noncontextual. \endproof

Then, consider a generic noncontextuality inequality of an $n$-cycle scenario. Since the inequality is linear, there must be at least one vertex of the $\mathcal{ND}$ polytope which maximally violates it. If there are two such vertices, then all convex combinations of them would also violate it maximally and, thus, be maximally contextual. However, according to the lemma proved above, an equally weighted convex combination of any two distinct such vertices is noncontextual \cite{MarcotulioDiss}. Thus, there can only be one vertex which maximally violates this inequality, and it also follows that this is the only behavior achieving this violation.

To complete the proof of Result \ref{result: max_contextual_implies_local}, we slightly modify an argument presented in Ref.~\cite{Masanes2006} to the generalized scenarios. Consider a behavior $p_{A_x, B_i B_{i+1}}$ such that Bob's marginal $p_{B_i B_{i+1}}$ is a vertex of his nondisturbing polytope. Then, we may write
\begin{equation}\label{eq:vertices_lead_to_uncorrelation}
    p_{B_i B_{i+1}}(b_i, b_{i+1}) = \sum_a p_{A_x, B_i B_{i+1}}(a, b_i, b_{i+1}) = \sum_a p_{A_x}(a)\ p^{A_x}_{B_i B_{i+1}}(b_i, b_{i+1}\vert a).
\end{equation}
Notice that, for each value of $a$, $p^{A_x}_{B_i B_{i+1}}(b_i, b_{i+1}\vert a)$ can be seen as a nondisturbing behavior of Bob's local experiment, and the above sum is a convex combination of such behaviors. However, since $p_{B_i B_{i+1}}$ is a vertex of Bob's nondisturbing polytope, then the above convex combination must be composed of only one term. That is, $p_{A_x, B_i B_{i+1}}(a, b_i, b_{i+1}) = p_{A_x}(a) p_{B_i B_{i+1}}(b_i, b_{i+1})$ and there is no correlation whatsoever between Alice and Bob.

\section{Maximal violation of generalized-CHSH inequalities implies noncontextuality}\label{app:example1}
In Ref.~\cite{xue2023}, the inequalities
\numparts
\begin{equation}\label{eq:CHSH63}
    \langle A_0 B_0 B_1 \rangle + \langle A_0 B_2 B_3 \rangle + \langle A_1 B_0 B_1 \rangle - \langle A_1 B_2 B_3 \rangle \leq 2
\end{equation}
and
\begin{equation}\label{eq:CHSH64}
    \langle A_0 B_0 \rangle + \langle A_0 B_2 B_3 \rangle + \langle A_1 B_0 \rangle - \langle A_1 B_2 B_3 \rangle \leq 2
\end{equation}
\endnumparts
were discovered to be generalized-Bell inequalities for generalized scenarios where Bob's local compatibility setup is either a $4$-cycle or a $5$-cycle. However, even for $n$-cycles with $n>5$ they are still satisfied by all local behaviors. This can be seen by noticing that the local maximum of such expressions is achieved in a local vertex of the associated $\mathcal{NSND}$ polytope, and all such vertices are product behaviors, that is, there isn't any correlation between Alice and Bob.

For example, let us analyze inequality \eref{eq:CHSH63} in a generalized scenario where Bob has any $n$-cycle. Considering the facts just mentioned, in a local vertex of the associated polytope, the inequality takes the form

\begin{equation}
    \langle A_0 \rangle \left ( \langle B_0 B_1 \rangle + \langle B_2 B_3 \rangle \right) + \langle A_1 \rangle \left( \langle B_0 B_1 \rangle - \langle B_2 B_3 \rangle \right)\leq 2.
\end{equation}
This can be straightforwardly verified by considering all possible values of $\langle A_0 \rangle, \langle A_1 \rangle, \langle B_0B_1 \rangle, \langle B_2 B_3 \rangle = \pm 1$ (Remember that all these measurements are dichotomic, with results labeled by $+1$ and $-1$).

On the other hand, regarding the maximal violations of inequalities \eref{eq:CHSH63} and \eref{eq:CHSH64} allowed by the no-signalling and no-disturbance conditions, by constructing analogues of PR-boxes one can verify that both of them can achieve their algebraic maximum $4$ within the $\mathcal{NSND}$ polytope.

Now, let us suppose that \eref{eq:CHSH63} is violated to its algebraic maximum. The only way in which this can be achieved is if $\langle A_0B_0B_{1}\rangle = \langle A_0B_2B_{3}\rangle = \langle A_1B_0B_{1}\rangle = -\langle A_1B_2B_{3}\rangle = 1$. 

This means that Alice's measurement outcomes and the product of Bob's measurement outcomes are perferctly correlated or perfectly anti-correlated, which implies that
\begin{equation}
    \langle A_0B_0B_{1}\rangle = 1 \Rightarrow  \langle A_0 \rangle =  \langle B_0B_{1}\rangle
\end{equation}
and
\begin{equation}
    \langle A_1B_2B_{3}\rangle = -1 \Rightarrow  \langle A_1 \rangle =  -\langle B_2B_{3}\rangle.
\end{equation}
These relations, in turn, lead to
\begin{equation}
    \langle A_0 \rangle  = \langle B_0 B_1 \rangle  = \langle A_1 \rangle    
\end{equation}
and
\begin{equation}
    \langle A_0 \rangle  = \langle B_2 B_3 \rangle  = -\langle A_1 \rangle.
\end{equation}
But these equations can only be satisfied if $ \langle A_0 \rangle  = \langle B_0 B_1 \rangle  = \langle A_1 \rangle = \langle B_2 B_3 \rangle = 0$. 

However, if Bob's contextuality scenario is an $n$-cycle, all noncontextuality inequalities are of the form
\begin{equation}\label{eq:cycle_nc_inequalities}
    \sum_{j=0}^{n-1} \gamma_j \langle B_j B_{j+1} \rangle \leq n-2
\end{equation}
where $\gamma_j = \pm 1$ and $\prod_j \gamma_j = -1$ \cite{Araujo2013}. By the result above, if inequality \eref{eq:CHSH63} is maximally violated, none of the noncontextuality inequalities of Bob's is violated, since two of the Bob's correlators $\langle B_j B_{j+1} \rangle$ are zero. Since these are all the noncontextuality inequalities of the $n$-cycle scenarios, this proves the noncontextuality of Bob's marginal behavior. A similar argument can be applied to \eref{eq:CHSH64}, and all the other CHSH-like inequalities obtained by considering other contexts of Bobs.

\section{Maximal violation of generalized-chained inequalities implies noncontextuality}\label{app:example2}
Consider a scenario where Bob realizes an $n$-cycle and Alice realizes at least $n$ incompatible measurements. It's possible to prove that the following generalization of \eref{eq:CHSH63} is a generalized-Bell inequality:
\begin{equation}\label{eq:Chained}
        \eqalign{\langle A_0 B_0B_1 \rangle +  \langle A_1 B_0B_1 \rangle +  \langle A_1 B_1B_2 \rangle + \dots \cr + \langle A_{n-1} B_{n-1} B_0 \rangle - \langle A_0 B_{n-1}B_0 \rangle \leq 2n-2.}
\end{equation}
Similarly to the inequalities discussed in the last section, one can prove that the maximum violation of \eref{eq:Chained} by $\mathcal{NSND}$ behaviors matches its algebraic maximum $2n$, and this can only happen if $\langle A_x B_jB_{j+1} \rangle = 1$ for all triple correlators appearing in the inequality, except for $\langle A_0 B_{n-1}B_0 \rangle = -1$. This implies that
\begin{equation}
    \langle A_0 \rangle = \langle B_0B_1 \rangle = \langle A_1 \rangle  =\dots = \langle B_{n-1}B_0 \rangle = - \langle A_0 \rangle.
\end{equation}
Thus, all the correlators $\langle A_i \rangle$ and $\langle B_j B_{j+1} \rangle$ must be zero, and, considering that all the noncontextuality inequalities of the $n$-cycle are of the form \eref{eq:cycle_nc_inequalities}, it follows Bob's marginal behavior is noncontextual.

\section{No trade-off in PM scenarios}\label{app:peres-mermin}
Consider the generalized Bell scenario where Alice has two incompatible measurements $A_0$ and $A_1$, and Bob has a PM square $\{B_{11}, ..., B_{33}\}$. In the following, we give an example of a $\mathcal{NSND}$ behavior which maximally violates the Peres-Mermin noncontextuality inequality
\begin{equation}\label{eq:PMinequality}
        \eqalign{\langle B_{11} B_{12} B_{31} \rangle + \langle B_{12} B_{22} B_{32} \rangle +\langle B_{13} B_{23} B_{33} \rangle + \cr + \langle B_{11} B_{12} B_{13} \rangle + \langle B_{21} B_{22} B_{23} \rangle - \langle B_{31} B_{32} B_{33} \rangle \leq 4}
\end{equation}
and the CHSH inequality
\begin{equation}\label{eq:standardCHSH}
    \langle A_0 B_{11} \rangle + \langle A_0 B_{21} \rangle + \langle A_1 B_{11} \rangle - \langle A_1 B_{21} \rangle \leq 4.
\end{equation}

Consider a behavior for which $\langle B_{11} B_{21} B_{31} \rangle = \langle B_{12} B_{22} B_{32} \rangle = \langle B_{13} B_{33} B_9 \rangle = \langle B_{11} B_{12} B_{13} \rangle = \langle B_{21} B_{22} B_{23} \rangle = 1$, $\langle B_{31} B_{32} B_{33} \rangle = -1$, thus maximally violating \eref{eq:PMinequality}. Moreover, suppose that it also has $\langle A_0 B_{11} \rangle = \langle A_0 B_{21} \rangle = \langle A_1 B_{11} \rangle = 1$ and $\langle A_1 B_{21} \rangle = -1$, thus maximally violating \eref{eq:standardCHSH}. To assure that this behavior is well defined, also consider that $\langle A_0 B_{21} B_{31} \rangle = \langle A_0 B_{12} B_{32} \rangle = \langle A_0 B_{12} B_{13} \rangle = \langle A_0 B_{21} B_{23} \rangle = \langle A_1 B_{21} B_{31} \rangle = \langle A_1 B_{12} B_{13} \rangle = 1$ and $\langle A_1 B_{12} B_{32} \rangle = \langle A_1 B_{21} B_{23} \rangle = -1$, and all the other correlators are zero. A lengthy, but straightforward calculation shows that the behavior generated by
\begin{equation}
        \eqalign{p_{A_x B_i B_j B_k}(ab_i b_j b_k) = \frac{1}{12}(1 + a\langle A_x \rangle + b_i \langle B_i \rangle + b_j \langle B_j \rangle 
        +b_k \langle B_k \rangle + \cr + ab_i \langle A_x B_i \rangle + ab_j \langle A_x B_j \rangle + ab_k \langle A_x B_k \rangle + \cr
        + ab_i b_j \langle A_x B_i B_j \rangle + ab_i b_k \langle A_x B_i B_k \rangle + \\
        + ab_j b_k \langle A_x B_j B_k \rangle + b_i b_j b_k \langle B_i B_j B_k\rangle)}    
\end{equation}
for all $x \in \{0,1\}$ and all valid contexts $\{i,j,k\}$ of the Peres-Mermin square \ref{fig:peres} respects positivity, normalization and all the no-signaling and no-disturbance conditions.

\end{document}